\documentclass[letterpaper, 10 pt, conference]{ieeeconf}  % Comment this line out if you need a4paper

\IEEEoverridecommandlockouts                              % This command is only needed if 
\usepackage{cite}
\usepackage{amsmath,amssymb,amsfonts}
\usepackage{algorithmic}
\usepackage{graphicx}
\usepackage{textcomp}
\usepackage{nicefrac}
\usepackage[table]{xcolor}
\usepackage{svg} % is this allowed?
\usepackage{comment}
\usepackage{lipsum}  
\usepackage{siunitx}
\usepackage{hyperref}

\title{\LARGE \bf
EEG Blink Artifacts Can Identify Read Music in Listening and Imagery
}

\author{
Abhinav Uppal$^1$, 
Dillan Cellier$^2$,
Min Suk Lee$^1$,
Sean Bauersfeld$^1$, 
Yuchen Xu$^3$,
Shihab A. Shamma$^4$,\\
Gert Cauwenberghs$^{1,3}$ and 
Virginia R. de Sa$^{2,3,5}$%
\thanks{
$^1$Shu Chien-Gene Lay Department of Bioengineering, 
$^2$Department of Cognitive Science,
$^5$Institute for Neural Computation, 
$^6$Hal\i c\i o\u glu Data Science Institute, 
UC San Diego, La Jolla, CA, USA.
}%
\thanks{$^{4}$Department of Electrical and Computer Engineering, Institute for Systems Research, University of Maryland, College Park, MD, USA.}%\thanks{\tt\small Contact: auppal@ucsd.edu}
\thanks{© 2025 IEEE. Personal use of this material is permitted. Permission from IEEE must be obtained for all other uses, in any current or future media, including reprinting/republishing this material for advertising or promotional purposes, creating new collective works, for resale or redistribution to servers or lists, or reuse of any copyrighted component of this work in other works.}
}

\def\BibTeX{{\rm B\kern-.05em{\sc i\kern-.025em b}\kern-.08em
    T\kern-.1667em\lower.7ex\hbox{E}\kern-.125emX}}

\begin{document}
\bstctlcite{IEEEexample:BSTcontrol}
\maketitle
\thispagestyle{empty}
\pagestyle{empty}

\begin{abstract}
Eye-movement related artifacts including blinks and saccades are significantly larger in amplitude than cortical activity as recorded by scalp electroencephalography (EEG), but are typically discarded in EEG studies focusing on cognitive mechanisms as explained by cortical source activity. 
Accumulating evidence however indicates that spontaneous eye blinks are not necessarily random, and can be modulated by attention and cognition beyond just physiological necessities. 
In this exploratory analysis we reanalyze a public EEG dataset of musicians listening to or imagining music (Bach chorales) while simultaneously reading from a sheet of music. 
We ask whether blink timing in reading music, accompanied by listening or imagery, is sufficient to uniquely identify the music being read from a given score. 
Intra-subject blink counts and timing are compared across trials using a spike train distance metric (Victor and Purpura, 1997). 
One trial-left-out cross-validation is used to identify the music being read with above chance level accuracy (best subject: 56\%, chance: 25\%), where accuracy is seen to vary with subject, condition, and a tunable cost factor for time shifts.
Future studies may consider incorporating eye blink contributions to brain decoding, especially in wearables where eye blinks could be easier to record than EEG given their higher amplitudes.
\end{abstract}

%\begin{IEEEkeywords}
%EEG, EOG, eye blinks, sight-reading, music %decoding, spike synchrony, Victor-Purpura distance
%\end{IEEEkeywords}

%\IEEEpeerreviewmaketitle

\section{Introduction}
\begin{figure}[ht]
\centerline{
\includegraphics[width=0.9\columnwidth]{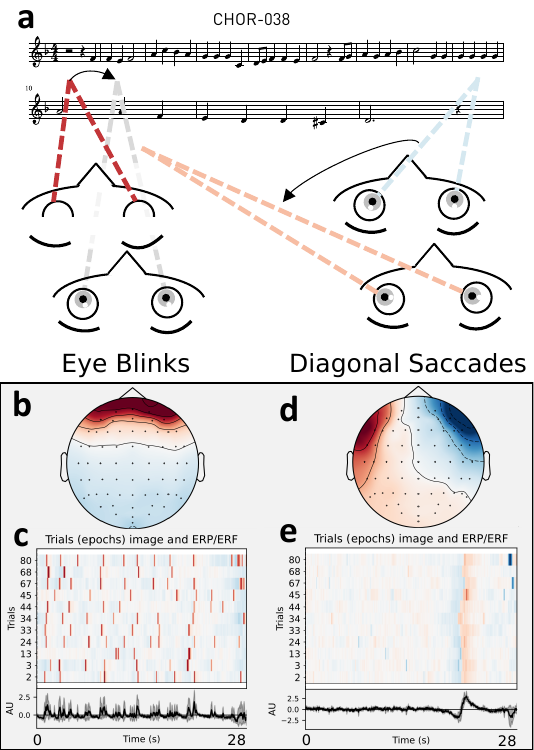}
}
\caption{
\textbf{a} One chorale (CHOR-038) from the music imagery dataset's sheet music, read by all subjects for 11 listening and 11 imagery trials per subject.
\textbf{b--c} Top independent component analysis (ICA) component corresponding to eye blinks for Subject 1, extracted from 11 listening trials of reading CHOR-038.
\textbf{d--e} Second ICA component for the same subject, chorale and condition (CHOR-038, listening) showing diagonal saccades (eye sweeps) between the two lines of music, occurring near the end of the first line of 9 bars, with expected duration $= 9 \times 2.4 s = 21.6 s$).}
\label{fig:ica}
\end{figure}

\begin{comment}
    - Music decoding in EEG
        - Artifacts in EEG
            - How blinks are identified and removed
    - Blinking is more than maintenance
        - When we blink
            - visual contrast
            - reading
                - text
                - music
        - When we don't blink
            - salience
            - cognitive load
            - response is required
            - synchronization in interaction?
        - Age/sex effects
        - Repeatability
            - F1 racing
            - Music performance from memory
    - Listening and imagery dataset
\end{comment}
Recent studies have investigated decoding music from non-invasive brain recordings using EEG 
\cite{
marionMusicSilencePart2021, 
libertoMusicSilencePart2021, 
dilibertoAccurateDecodingImagined2021, 
chungDecodingImaginedMusical2023},
joint EEG-fMRI \cite{dalyNeuralDecodingMusic2023}, 
or EEG-fNIRS \cite{zhuDesignImplementationEEGfNIRS2024}.
For example, \cite{marionMusicSilencePart2021, dilibertoAccurateDecodingImagined2021} successfully decode heard or imagined music from brain activity of musicians by recording their EEG. 
But as EEG is susceptible to interference from non-brain sources such as line noise (50 Hz or 60 Hz), eye movements, muscle activity (EMG), and motion artifacts, such artifacts are often isolated and removed from EEG analyses to improve the signal-to-noise ratio of underlying cortical activity.
Specifically, eye-blink related artifacts in EEG are removed by either dropping data segments with ocular artifacts, or by decomposing and removing eye-related components using source-separation methods such as Independent Component Analysis (ICA, see Fig.\ref{fig:ica}), or with joint eye-tracking methods \cite{plochlCombiningEEGEye2012}.

% Blinks: more than lubrication
Eye movements, although artifactual in EEG analyses, can carry task-relevant information and are implicated in perception \cite{
%finkMusicModulatesEyeblinks2014, 
langeEyeBlinkingMusical2023,
willettPerceptualConsequencesNeurophysiology2023}. 
Spontaneous eye blinks can occur every few seconds, which is more frequent than required for physiological necessities such as lubricating the eyes or maintaining the tear-film \cite{
%nakanoBlinkrelatedMomentaryActivation2013,
yangEyeBlinksVisual2024}. 
While blinking can enhance visual contrast \cite{yangEyeBlinksVisual2024}, 
blinking can also be inhibited when salient stimuli are anticipated \cite{shultzInhibitionEyeBlinking2011}, 
during cognitive loading \cite{siegleBlinkYouThink2008}, 
or in allocation of auditory attention when responding to stimuli is required \cite{huberPatternsEyeBlinks2022}.
Blinking is not just related to audiovisual stimuli onsets or preparation for motor action, but blink modulation can also be attributed to top-down cognitive processing \cite{muraliLatencySpontaneousEye2021}.
Finally, blink timing is repeatable within individuals, e.g. in F-1 race-car driving practice on a concourse \cite{nishizonoHighlyReproducibleEyeblink2023}, 
or even when performing piano music from memory \cite{finkEyeMovementPatterns2023}. %as needed for a biomarker. 
Therefore, eye blinks could be leveraged by brain-machine interfaces in circumstances where high quality EEG data may not be accessible. 

In this work, we extract eye blinks from a public EEG dataset for music listening and imagery \cite{marionDataMusicSilence2021}, and ask whether blink-timing is sufficient to identify the music being read by trained musicians from a given score (Fig. \ref{fig:2}a).
We chose this dataset as it demonstrably relates EEG responses to melodic expectation or surprise\cite{marionMusicSilencePart2021}, leading us to hypothesize corresponding modulation of eye blinks, given that blinks are implicated in cognitive processing, sight-reading music \cite{finkMusicModulatesEyeblinks2014}, and stimulus predictability \cite{bonnehBlinkingSurpriseEyeBlink2015}.

The rest of this report describes the dataset in Sec. \ref{sec:dataset}, eye-blink extraction in Sec. \ref{sec:blink}, a spike-timing metric adapted from neuronal analysis for comparing blink times in Sec. \ref{sec:distance}, and our music decoding results and discussion in Sec. \ref{sec:results} and Sec. \ref{sec:discussion}, respectively.

\section{Methods}
\subsection{Public Music Imagery Dataset}
\label{sec:dataset}
\begin{figure}[t]
\centerline{\includegraphics[width=\columnwidth]{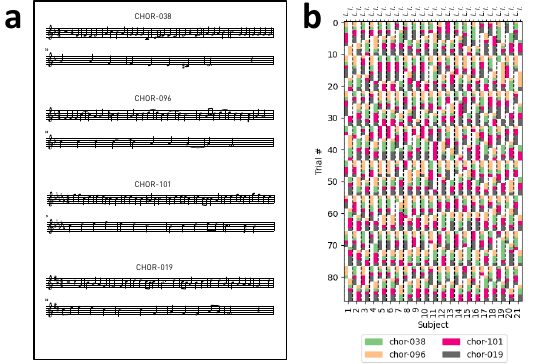}}
\caption{
\textbf{a} Sheet music \texttt{Score.pdf} from the Music Imagery Dataset \cite{marionDataMusicSilence2021}, consisting of four monophonic Bach chorales (CHOR) with 12 bars (CHOR-038, 096, 101) or 13 bars (CHOR-019) spanning two rows each.
CHOR-101 has 8 bars in its first row, while the rest have 9 bars on top.
\textbf{b} Experiment design matrix showing the trial ordering of the four chorales for both conditions ($L:$ Listening, $I:$ Imagery) as two columns per subject, for all $N=21$ subjects. One chorale was read per trial in either condition with a vibrotactile metronome on the left ankle for maintaining timing.
}
\label{fig:2}
\end{figure}

A public EEG dataset of $N=21$ expert musicians listening to or imagining four Bach chorales was obtained from \cite{marionDataMusicSilence2021, CNSPInitiative}. 
The experiment consisted of trials where subjects read one of four chorales from the sheet music shown in Fig. \ref{fig:2}a, while imagining or listening to the music at 100 beats per minute (BPM) for about \SI{30}{\second}.
11 such trials were recorded for each of four chorales and two conditions (listening or imagery), resulting in 88 trials per subject. 
Fig. \ref{fig:2}b shows the trial order of conditions and chorales for each subject, showing randomization while grouping listening and imagery trials into blocks. 

Subjects were intimately familiar with the music and rehearsed the pieces before EEG data collection.
The sheet music itself was affixed to a table in a dimly lit sound booth where data was collected. 
Imagery timing was maintained via a vibrotactile metronome attached to the left ankle, which cued the subjects at every bar onset (\SI{2.4}{\second}) for all imagery and also listening trials. 
More information about the dataset is available in Table \ref{tab:dataset}, with EEG decoding results with artifact removal in \cite{
marionDataMusicSilence2021, 
libertoMusicSilencePart2021, 
dilibertoAccurateDecodingImagined2021}.

\begin{table}[ht]
\centering
\caption{Music Imagery Dataset Parameters}
\begin{tabular}{|l|l|}
     \hline
     \textbf{Parameter} & \textbf{Value} \\
     \hline
     Subjects & 21 Musicians (1--8 analyzed)\\
     \hline
     Age & Median$=$25\\
     \hline
     Conditions & 2: Listening, Imagery\\
     \hline
     Sheets of Music (Score) & 1 (Single-Sided)\\
     \hline
     Number of Chorales (Songs) & 4 \\
     \hline
     Composer & Johann Sebastian Bach \\
     \hline
     Keys & Original \\
     \hline
     Sound (Listening Trials) & Monophonic (Fender Rhodes)\\
     \hline
     Beats Per Minute (BPM) & 100 \\
     \hline
     Bars per Chorale (1 Trial) & 12 \\
     \hline
     Duration of 1 Bar & 2.4 seconds (4 $\times$ \nicefrac{1}{4} notes)\\
     \hline
     Usable Duration$^{\mathrm{a}}$ of 1 Trial & 12 $\times$ 2.4 $\approx$ 28 seconds \\
     \hline
     Cue Modality (All Trials) & Tactile (Left Ankle) \\
     \hline
     Cue Onset & Bar (Every 2.4 seconds) \\
     \hline
     Number of Listening Trials & 4 chorales $\times$ 11 runs $=$ 44 \\
     \hline
     Number of Imagery Trials & 4 chorales $\times$ 11 runs $=$ 44 \\
     \hline
     Total Number of Trials & 88 per Subject\\
     \hline
     EEG System & BioSemi Active II\\
     \hline
     Number of Electrodes & 64 \\
     \hline
     Sample Rate & Band-pass filter: \SI{0.1}{\hertz}--\SI{30}{\hertz}\\ 
      & Downsampled to \SI{64}{\hertz} \\
     \hline
     \multicolumn{2}{l}{$^{\mathrm{a}}$of $\approx$35 seconds including note reference and cueing beats}
     \end{tabular}
\label{tab:dataset}
\end{table}

\subsection{Detection and Annotation of Eye Blinks}
\label{sec:blink}
\begin{figure}[ht]
    \centering
    \includegraphics[width=0.95\columnwidth]{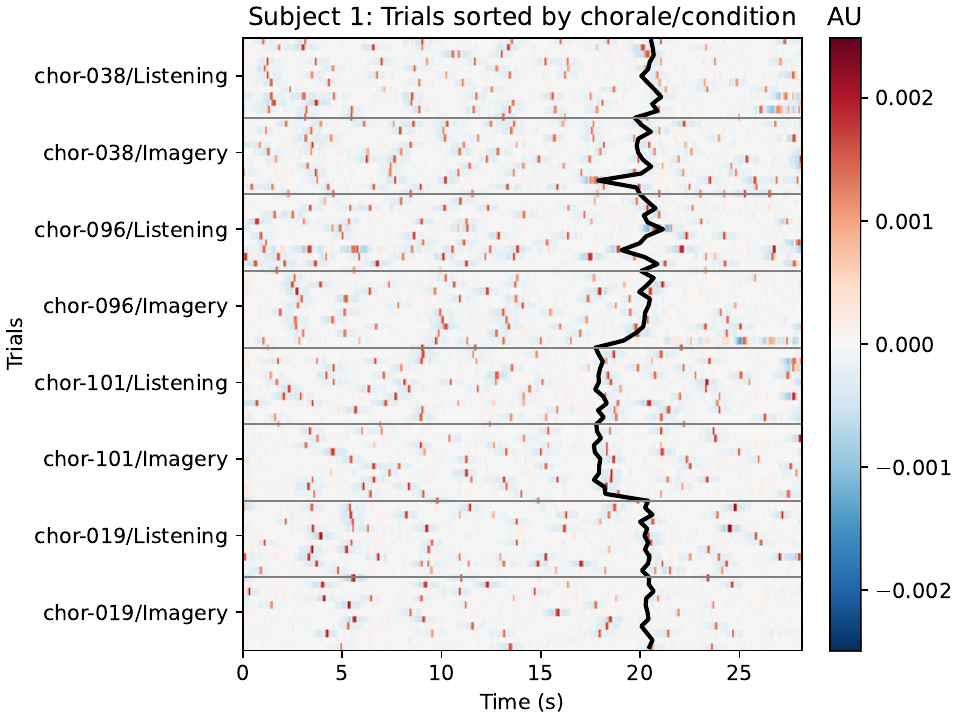}
    \caption{
    Independent component (IC) showing eye blinks (red) for Subject 1 over all 88 trials, grouped by chorale (chor) and condition (Listening or Imagery), hence not in original experiment order.
    IC amplitude is in arbitrary units (AU).
    Jagged thick black overlay indicates timing of sweeping eye motions (diagonal saccades) as the subject progressed from the first row of music to the second, with all CHOR-101 trials showing earlier sweeps corresponding to fewer (8) bars in the its first row of sheet music.}
    \label{fig:sub1-blinks}
\end{figure}

The ontained EEG dataset was already segmented into trials and pre-processed (Table \ref{tab:dataset}), with breaks in between trials removed.
The segmented dataset was imported and merged using EEGLAB \cite{delormeEEGLABOpenSource2004}.
Independent component analysis (ICA) and ICLabel \cite{pion-tonachiniICLabelAutomatedElectroencephalographic2019} were run to identify eye-related independent components (ICs).
Eye-related ICs were visually inspected and categorized as blinks or diagonal saccades (Fig.\ref{fig:ica}).
As not all subjects blinked during all trials, diagonal saccade ICs were inspected to verify the timing of eye-sweeps corresponding to the end of the  first row of sheet music for each chorale (Fig. \ref{fig:sub1-blinks}). 
All subjects showed diagonal saccades near times corresponding to the end of the first line, confirming all subjects read from the sheet music synchronous with the vibrotactile metronome.
Visualizations of IC components were made using MNE-Python \cite{gramfortMEGEEGData2013}.

Next, BLINKER \cite{kleifgesBLINKERAutomatedExtraction2017} was run to extract eye-blink times from the blink-related ICs using heuristically selected parameters for each subject, shown in Table \ref{tab:counts}.
For the 41 minutes of data available per subject, eye blinks identified by BLINKER were examined and manually corrected in cases where blinks were missed, or saccades had been identified as blinks.
Some subjects showed series of rapid blinks. 
Instances of overlapping blinks and saccades were also seen.
As no ground truth was available for blink times in the dataset, manual judgment was used to correct blink annotations.
Given the subjective nature of these corrections, only musicians 1--8 were analyzed for this pilot exploration of blink timing based music decoding.
Subjects 2 and 5 had some trials with no detected blinks and were thus dropped from further analysis.
Time-series of blink times (blink trains) were exported per subject, condition, and trial for further analysis.

\subsection{Victor-Purpura Spike Train Distance}
\label{sec:distance}
%Choice of distance metric. 
Time-series of spike-times (spike trains) have been extensively analyzed in the neuroscience literature, and multiple methods are available for computing the distance (or dissimilarity) of two spike trains.
As a metric for spike-train dissimilarity is generalizable to pair-wise comparisons of event sequences, eye-tracking studies have utilized neuronal spike-train analysis methods such as the Victor-Purpura distance \cite{victorMetricspaceAnalysisSpike1997} and spike-synchrony measures \cite{kreuzMonitoringSpikeTrain2013}.
Motivated from a previous study analyzing eye blink trains as neuronal spike trains \cite{nomuraEmotionallyExcitedEyeblinkrate2015}, we applied 
the Victor-Purpura spike train distance \cite{victorMetricspaceAnalysisSpike1997} $D_{spike}(q)$, such that spike-times correspond to blink times here, and $q$ (\SI{}{\hertz}) is a hyper-parameter for penalizing time-shifts between blinks across trials.

To compare intra-subject blink times between two trials $A$ and $B$, the Victor-Purpura distance can be used to assign a minimum cost for transforming each blink time from trial $A$ to $B$ as: 
\begin{equation}
    Cost = \begin{cases}
        1: \text{Add a blink}\\
        1: \text{Remove a blink}\\
        \frac{\Delta T}{\nicefrac{1}{q}}: \text{Time-shift a blink by time } \Delta T
    \end{cases}
\end{equation}

Intuitively, we think of \nicefrac{1}{q} as parameterizing time-shift invariance in the cost of shifting blink times. 
To consider two extremes, \nicefrac{1}{q}$=$\SI{0}{\second} (equivalently $q=\infty\,$\SI{}{\hertz}) penalizes any shifting of blink times, such that every dissimilar blink time between trials $A$ and $B$ costs 1 unit.
On the other hand, setting \nicefrac{1}{q}$=\infty\,$\SI{}{\second} (or $q=0$ \SI{}{\hertz}) allows free translation of blink times, such that the only cost comes from mismatched number of blinks between trials $A$ and $B$.

We calculated intra-subject Victor-Purpura distance for all pair-wise trials separately for the two conditions, music listening and imagery, stepping \nicefrac{1}{q} from \SI{0}{\second} to the trial duration, in steps corresponding to each beat (\SI{600}{\milli\second}), and also for \nicefrac{1}{q}$=\infty$ \SI{}{\second}. 
The implementation of Victor-Purpura distance from Elephant was used \cite{elephant18}. 

\subsection{Intra-subject Decoding of Read Music}
%Leave-one-out cross-validation.
Leave-one-trial-out cross-validation was used to identify the music being read by each subject, separately in music listening and imagery conditions.
One of the 44 trials per subject per condition (listening or imagery of 4 chorales) was left-out. 
Victor-Purpura distance of blink times from this left-out trial was calculated with each of the remaining 43 trials from the same condition, spanning all four chorales.
The chorale label of the closest trial (smallest distance) was selected as the decoded music label. 
This process was repeated for leaving out each of the 44 trials per condition.
Decoding accuracy was then determined as the number of 44 left-out trials that were identified correctly.
Confusion matrices were generated to determine which chorales tended to get misclassified.
For each left-out trial, chance-level probability for choosing a matching chorale (10 trials) among the 43 trials was \nicefrac{10}{43}, approximately 25\%. 

\subsection{Code Availability}
All code associated with this work is available on \href{https://github.com/3x10e8/eye-blink-music}{https://github.com/3x10e8/eye-blink-music}.

\section{Results}
\label{sec:results}

\subsection{Total Blink Counts}
Table \ref{tab:counts} shows BLINKER parameters used for each of 8 subjects,  total number of ``good" blinks detected by BLINKER, and final blinks counts after manual annotation (Annot.).
Blinking rates (number of blinks per minute) were seen to vary across subjects.

\begin{table}[b]
    \centering
    \caption{Total detected blinks across all 88 trials (approx. 41 minutes per subject)}
\begin{tabular}{|c|c|c|c|c|c|c|}
\hline
 & \multicolumn{4}{c|}{BLINKER Parameters and Results} & \multicolumn{2}{c|}{Manual} \\
 \cline{2-7}
\textbf{Sub-} & \textbf{Filter$^\mathrm{a}$} & \textbf{Stdev.$^\mathrm{b}$} & \textbf{Corr.$^\mathrm{c}$} & \textbf{Good} &  \textbf{Annot.} & \textbf{Rate}\\
\textbf{ject} & [Hz--Hz] & Thresh. & Thresh. & Blinks & Blinks & [/min] \\
  \hline
1 & 1--20 & 0.2 & 0.9 & 681 & 614 & 15\\
\hline
2 & \multicolumn{6}{c|}{not enough eye blinks}\\
\hline
3 & 1--20 & 1.5 & 0.9 & 334 & 363 & 9 \\
4 & 1--20 & 2.5 & 0.9 & 294 & 309 & 7 \\
\hline
5 & \multicolumn{6}{c|}{not enough eye blinks}\\
\hline
6 & 2--20 & 2.2 & 0.2 & 536 & 396 & 10 \\
7 & 2--20 & 1.5 & 0.9 & 1518 & 1575 & 38 \\
8 & 1--20 & 6.0 & 0.95 & 272 & 284 & 7 \\
        \hline
    \multicolumn{7}{l}{BLINKER parameters $^\mathrm{a}$lowCutoffHz--highCutoffHz, $^\mathrm{b}$stdThreshold, and}\\
    \multicolumn{7}{l}{$^\mathrm{c}$correlationThresholdBottom.}\\
    \end{tabular}
    \label{tab:counts}
\end{table}

\subsection{Decoding Read Music in Listening and Imagery}
\begin{figure}
    \centering
    \includegraphics[width=0.95\columnwidth]{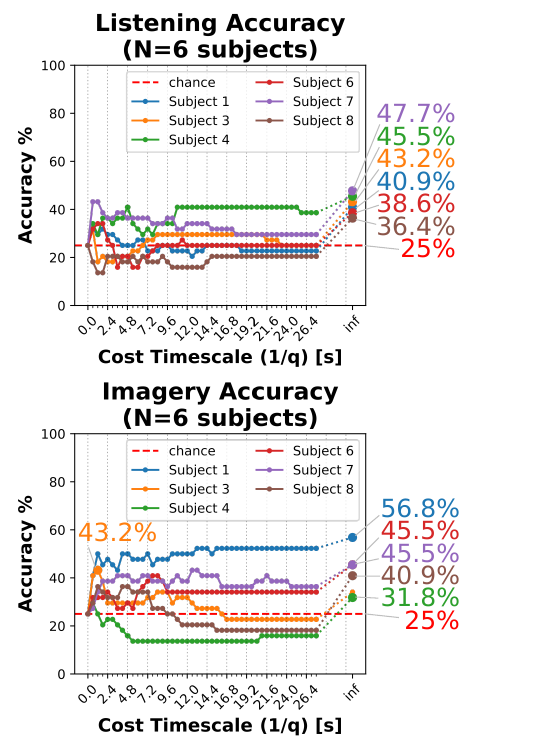}
    \caption{Intra-subject sight-read music classification accuracies using Victor-Purpura distance between blink times with one-trial left-out cross validation. Hyperparameter $q$ (cost factor) was swept for each subject, with highest accuracies seen for $q=0$\SI{}{\hertz} (equivalently \nicefrac{1}{q}$=\infty\,$\SI{}{\second}), except for Subject 3 in the imagery condition.}
    \label{fig:accuracy}
\end{figure}

\begin{figure}
    \centering
    \includegraphics[width=0.95\columnwidth]{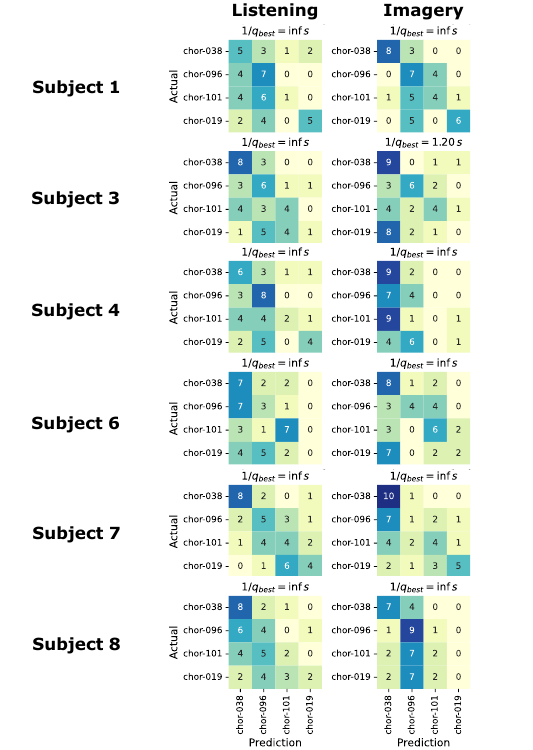}
    \caption{Confusion matrices for classifying sight-read music for each subject and listening or imagery conditions, corresponding to the best accuracy hyperparamter from Fig.\ref{fig:accuracy}, here labelled as $q_{best}$.}
    \label{fig:confusion}
\end{figure}

Fig. \ref{fig:accuracy} shows accuracies for decoding the read chorales using Victor-Purpura distance of left-out trials, separately for listening and imagery conditions per subject.
Decoding accuracies are shown for time-steps of the cost timescale \nicefrac{1}{q} corresponding to a beat duration of \SI{600}{\milli\second}, ending with \nicefrac{1}{q}$=\infty$ \SI{}{\second} corresponding to no penalty for shifting blink times between compared trials. Confusion matrices corresponding to the best accuracy $q$, labeled $q_{best}$, are shown in Fig. \ref{fig:confusion} for each subject and condition.

\section{Discussion \& Conclusion}
\label{sec:discussion}
This work examined decoding of sight-read music under two conditions with possibly different cognitive loading: music listening and imagery, as performed by expert musicians in a public EEG dataset. 
Not all subjects sufficiently blinked across sight-reading trials (2 of 8 examined), and were dropped from analysis, highlighting a limitation that blink inhibition in some subjects could limit the usability of blink-based decoding.
For subjects that did show blinking activity across multiple trials, the read music could be identified correctly for all subjects and both listening and imagery conditions with above chance-level accuracy.
Best subjects showed decoding accuracies of 47.7\% in music listening (Subject 7) and 56.8\% in music imagery (Subject 1). In both cases, maximum accuracies were seen when only total blink counts were considered, corresponding to a Victor-Purpura cost factor $q=0$ \SI{}{\hertz} (equivalently \nicefrac{1}{q} $=\infty$ \SI{}{s}).
This highlights a second limitation of blink-based decoding, where only activity over longer timescales could be suitable for blink-based decoding such that enough blinks could have occurred under cognitive loading. 
One subject, however, did show highest accuracy for imagery decoding on a shorter timescale of two bars (\nicefrac{1}{q} $=$\SI{1.2}{\second}).

As the utilized public dataset was from an EEG study focusing on brain activity, ground-truth eye-movements were not available for validating BLINKER and manual annotations of blinks as utilized in this work.
Follow-up datasets may consider joint EEG and eye-tracking methods when available and not disruptive to the task.
Additionally, re-analysis of existing EEG datasets could benefit from using artificial neural network approaches such as EEGEyeNet \cite{kastratiEEGEyeNetSimultaneousElectroencephalography2021a} to complement blink detection from single-channel (or single independent component) based methods, as utilized here, to improve blink detection by considering multiple channels.
%Eye blinks are inhibited in reading \cite{chidi-egbokaBlinkRateMeasured2023}. 
%Gaze angle confound \cite{doughtySpontaneousEyeblinkActivity2014}.

For a metric for blink timing comparison, future studies may also consider and contrast other spike-train synchrony measures such as 
SPIKE-distance \cite{kreuzMonitoringSpikeTrain2013}, as used for comparing blink times in an F-1 race-car driving study \cite{nishizonoHighlyReproducibleEyeblink2023}.
Finally, analysis of EEG data recorded from wearables could be considered, to evaluate whether including eye blinks in addition to wearable EEG could enhance decoding performance, compared to only using EEG with ocular artifacts discarded. 

In summary, we provide support for the case of eye blinks being task-relevant, where blink timing is modulated repeatably within subjects to allow for above chance-level decoding of read music in expert musicians.
Data constrained settings such as wearables could thus benefit from decoding user attention and cognitive loading with blinks, with the possibility to supplement EEG-based performance, instead of discarding blink data completely as artifactual.

%\addtolength{\textheight}{-25cm}

\section{Acknowledgment}
We thank G. Marion, 
G. M. Di Liberto, 
C. Pelofi, 
M. Rezaeizadeh, 
A. Karsolia, 
L. Fink, 
B. Voytek, 
participants of the 2022 Telluride Neuromorphic Cognition Engineering Workshop, and 
NSF IIS 1817226.

\bibliographystyle{IEEEtran}
\bibliography{root-preprint}
\end{document}